\documentclass[aip,pop,preprint]{revtex4-1}
\usepackage{graphicx}
\usepackage{dcolumn}
\usepackage{bm}
\usepackage{color}
\usepackage{tikz}
\begin{document}


\title{Evolution of ultra-relativistic hollow-electron-beam wakefield
drivers  during their propagation in plasmas} 


\author{Neeraj Jain}
\affiliation{Zentrum f\"ur Astronomie und Astrophysik, Technische
  Universit\"at Berlin, Hardenbergstr. 36, D-10623, Berlin, Germany}
\altaffiliation{Max Planck Institute for Solar System Research,
Justus-von-Liebig-Weg 3, 37077, G\"ottingen, Germany}



\date{\today}

\begin{abstract}
Ultra-relativistic hollow electron beams can drive plasma wakefields
($\sim$ GV/m)
suitable for  positron acceleration. 
Stable propagation of hollow electron beams for long distances in plasmas
is required to accelerate positrons to high energies by these plasma wakefields.
In this work,  we show by quasi-static kinetic simulations using the code WAKE 
that an ultra-relativistic  azimuthally-symmetric hollow 
electron beam propagates in a plasma by developing fish-bone like
structure and shifting its bulk, differentially along its
length (rear part
fastest), towards its axis due to the decrease in the betatron time period
of beam electrons from the beam-front to beam-rear.  
Hollow electron beams with small radius collapse into their axis due to the pull
  by the secondary wakefields generated by some of the beam electrons
reaching the axis.
Hollow beams with sufficiently large radius, however, can propagate
stably in plasmas for several meters and be used for positron acceleration.  
\end{abstract}

\pacs{}

\maketitle

\section{Introduction \label{sec:introduction}}

Plasma based
particle acceleration schemes, first proposed in 1979 \cite{tajima1979}, 
employ either a high
intensity laser or an ultra-relativistic charged particle beam driver which
propagates through a plasma driving electromagnetic fields in its 
wake where charged particles can be accelerated. 
In the blow-out regime of plasma wake field acceleration \cite{rosenzweig1991,lu2006}, the head of a short 
electron beam driver expels plasma
electrons radially away from its path setting up nonlinear plasma
oscillations. The expelled electrons fall back to and cross the beam
axis in a tightly 
confined region behind the beam
driver. As a result, the beam driver travels in a plasma-electron-free
region which is a bubble-like structure in the moving frame of the
beam. A favorable region (longitudinal field
accelerating and transverse field focusing) for electron acceleration  forms behind the
beam driver and inside the first bubble containing the driver. 
Experiments have
demonstrated multi-GeV energy gain of electrons accelerated    
by plasma wakefields driven by an ultra-relativistic electron beam in the
blow-out regime 
\cite{hogan2005,blumenfeld2007,litos2014,corde2016,joshi2018} and also by proton
beam \cite{adli2018}.  However,
acceleration of positrons, which is essential for a successful
operation of an electron-positron collider, using plasmas is relatively less
explored.

Current schemes of positron acceleration 
drive plasma wake fields  
either by electron \cite{lotov2007,wang2008,jain2015a} or
 by  positron beam drivers \cite{lee2001,blue2003,kimura2011,corde2015,doche2017}. 
In the blow-out regime driven by 
electron beam drivers,  a narrow 
favorable region for positron acceleration   forms between the first
and second bubbles.  
In the favorable region for positron acceleration, accelerating  electric field varies rapidly with  axial
coordinate leading to a large energy spread in the
accelerated witness beam of positrons. An efficient acceleration of
positrons in this scheme is possible only if the energy
content of the nonlinear plasma waves driven by the electron beam is small \cite{lotov2007}.
In the case of a positron beam driver, plasma electrons are attracted towards
rather
than expelled out of the beam driver path and do not cross the axis in
a confined 
region as they do in the case of an  electron beam driver \cite{lee2001}. As a result
the accelerating fields are smaller than those driven by electron beam drivers. 
The focusing fields are radially nonlinear and vary along the beam axis
 leading to an emittance growth of the witness positron beam
 \cite{muggli2008}. Strength of the
accelerating fields and quality of the focusing forces are improved
if the positron beam propagates through a hollow plasma channel
\cite{lee2001,kimura2011,yi2014}. Misalignment of positron beam from
the axis of the hollow plasma channel, however, deflects the beam and
sets stringent conditions on the beam alignment \cite{lindstrom2018}. 

A novel scheme using ultra-relativistic hollow
electron beams to drive plasma wakefields for positron acceleration was proposed \cite{jain2015a}.
In this scheme,  an annular shaped electron free region (bubble) forms 
generating the wakefields for positron acceleration in the hollow region
near the axis of the hollow beam. The accelerating field for positrons
in this favorable region of positron acceleration increases with
the total charge in the beam driver while the axial size of the
favorable region remains approximately of the order of a plasma
wavelength which is large enough for the placement of a witness beam
of positrons. This is in
contrast to the case of a solid beam driver in which the size of the favorable
region diminishes with increasing charge in the beam driver. 
Plasma wakefields for positron acceleration similar to those driven by
a hollow electron beam  can also be driven by 
Laguerre-Gaussian laser pulses \cite{vieira2014}. 

Hollow electron beams with energies in the range from tens to hundreds
of MeV have been observed to form in experiments and
simulations of laser wake field acceleration
\cite{pollock2015,zhao2016,zhang2016a,zhang2016b,shen2017}. These
beams form as a result of the trapping of the self or externally
injected electrons in the laser driven plasma wake fields and are of
interest for their applications to compact radiation sources. 
For applications to 
positron acceleration to high energies, ultra-relativistic
hollow electron beam wakefield drivers  with large energy contents, of the order of
tens of GeV and larger, are of particular interest. Their stable
propagation in plasmas for long distances is desirable for the success
of their application to positron acceleration.  
  Azimuthally symmetric 2-D
simulations demonstrated stable propagation of a hollow electron beam
with 23 GeV energy 
 in a plasma for certain plasma and beam parameters, accelerating a
 positron beam  (total charge 13.58
pC) to gain 12.4 GeV in 140 cm 
\cite{jain2015a}. Stable propagation of hollow electron beam, however,
depends on the parameters (plasma density, beam radius, thickness and
charge) which are subjected to variation under optimization of future
experiments.  In this paper
we study evolution of an ultra-relativistic hollow electron beam
driver propagating in a plasma  
and its dependence on the beam radius by quasi-static simulations. We find
that a hollow electron beam propagates in a plasma by developing
fish-bone like structure and shifting its bulk towards its
axis. Hollow beams
with small radius 
collapse into their axis while those with  sufficiently large radius can
propagate stably in plasmas.


The paper is organized as follows. Section \ref{sec:simulation}
presents simulation setup for our studies. Evolution of hollow
electron beams is discussed in Section \ref{sec:evolution} and its
dependence on hollow beam radius in Section \ref{sec:dependence_r0}.
Finally conclusion are presented in section \ref{sec:conclusion}.
\section{2-D Quasi-static simulations \label{sec:simulation}}
We simulate propagation of hollow
electron beams in uniform plasmas in an azimuthally symmetric
($\partial/\partial\theta=0$) cylindrical geometry ($r,\theta,z$) using the quasi-static code
WAKE
\cite{mora1997,jain2015b}.  Quasi-static approximation exploits the disparity of the
time scales of evolution of plasma and beam driver. 
The time scale of evolution of an ultra-relativistic electron beam driver
(relativistic factor $\gamma_b >> 1$) is the betatron period
$\tau_{b}=\sqrt{2\gamma_b}\lambda_p/c$ which is
much larger than the plasma time scale $\lambda_p/c$, where
$\lambda_p=2\pi c/\omega_p$ is the
plasma wavelength, $c$ is the speed of light and $\omega_p$ is the
electron plasma frequency. In the code WAKE,
fast response of the kinetic, warm and
relativistic plasma to the beam driver is calculated by solving quasi-static equations \cite{jain2015b}
for the wakefields and plasma electrons assuming a non-evolving beam
driver in a
computational domain which changes its position in the direction of
the beam propagation with the beam. The beam 
driver is then evolved  under the influence of the plasma wakefields
by solving the equations of motion for beam particles over longer time
scales.

We use transverse Coulomb gauge under which azimuthally symmetric
electromagnetic fields  are described by the electrostatic potential $\phi$ and
vector potential $\mathbf{A}=(0,0,A_z)$. Quasi-static equations  \cite{jain2015b}
for plasma electrons and wakefields can be derived by making the mathematical
transformations, $\xi=ct-z$ and  $s=z$, and  the approximation $\partial/\partial s <<
\partial/\partial \xi$. Under quasi-static approximation, axial and radial wakefields, $E_z$ and
$E_{rad}=E_r-cB_{\theta}$, respectively, can be obtained from the
wakefield potential $\psi=\phi-c A_z$ as $E_z=\partial\psi/\partial\xi$ and
$E_{rad}=-\partial\psi/\partial r$.

We take initial number density of the hollow electron beam driver as,
\begin{eqnarray}
 n_{b}&=&n_{b0}
\exp\left[-\frac{(r-r_0)^2}{2\sigma_r^2}-\frac{(\xi-\xi_0)^2}{2\sigma_z^2}
\right ]\label{eq:driver}
\end{eqnarray}
The peak number density ($n_{b0}$) of the hollow beam is located at 
an off axis location $(r_0,\xi_0)$ and falls off within distances $\sigma_r$
(radially) and $\sigma_z$ (axially). The beam is completely hollow in the limit
$r_0/\sigma_r\rightarrow \infty$. Otherwise there is a small but finite density
in the core of the beam. In the limit $r_0\rightarrow 0$, the beam density has
peak at the axis and corresponds to a solid beam.

The background plasma has a uniform 
density, $n_{p0}=5\times 10^{16}$ cm$^{-3}$ giving $k_p^{-1}=\omega_p/c=23.79 \,\mu$m, and is
modeled using 9 simulation particles per cell. The parameters for electron beam
driver are $n_{b0}=n_{p0}$, $\xi_0=0$, $\sigma_z=23.79 \,\mu$m $=k_p^{-1}$, $\sigma_r=4.76
\,\mu$m $=0.2 k_p^{-1}$. We shall vary the hollow beam radius $r_0$. The total
charge $Q_d=-e\int n_{b}(r,\xi) r\,dr\,d\theta\,d\xi$ contained in the beam driver
will thus vary with $r_0$. The driver beam has initial energy
of 23 GeV and is modeled using 1.25 $\times 10^6$ simulation particles.  The simulation domain size along
$\xi$ is approximately $262 \,\mu$m $\approx 11 \,k_p^{-1}$ with a grid
resolution $d\xi=0.52 \,\mu$m $\approx 0.02 \,k_p^{-1}$. The simulation domain size and
grid resolution in radial direction are approximately $119\,\mu$m
$\approx 5\,k_p^{-1}$ and
$dr\approx 0.3 \,\mu$m $\approx 0.0125\,k_p^{-1}$, respectively. The driver beam is propagated in the
uniform plasma in small steps of propagation distance $ds\approx 4.75
\,\mu$m. 

\section{Evolution of hollow electron beams \label{sec:evolution}}
Figs. \ref{fig:rhop_evolution_r08sr} and
\ref{fig:erad_evolution_r08sr} show evolution of an ultra-relativistic hollow
electron beam driver with radius $r_0=8\,\sigma_r=1.6\,k_p^{-1}$. The hollow beam develops fish-bone
like structure, shifts its bulk towards its axis and finally collapses
into its axis  during its
propagation in
a plasma under the influence of the plasma wakefields it
drives. Plasma wakefields driven by the initial beam density play an 
important role in the evolution of the beam and thus we discuss them first.
\subsection{Plasma wake-fields driven by the initial hollow beam density\label{sec:initial}}
\begin{figure}
\includegraphics[clip=true,trim=1.55cm 5.8cm 3.45cm
  2.5cm,width=\linewidth]{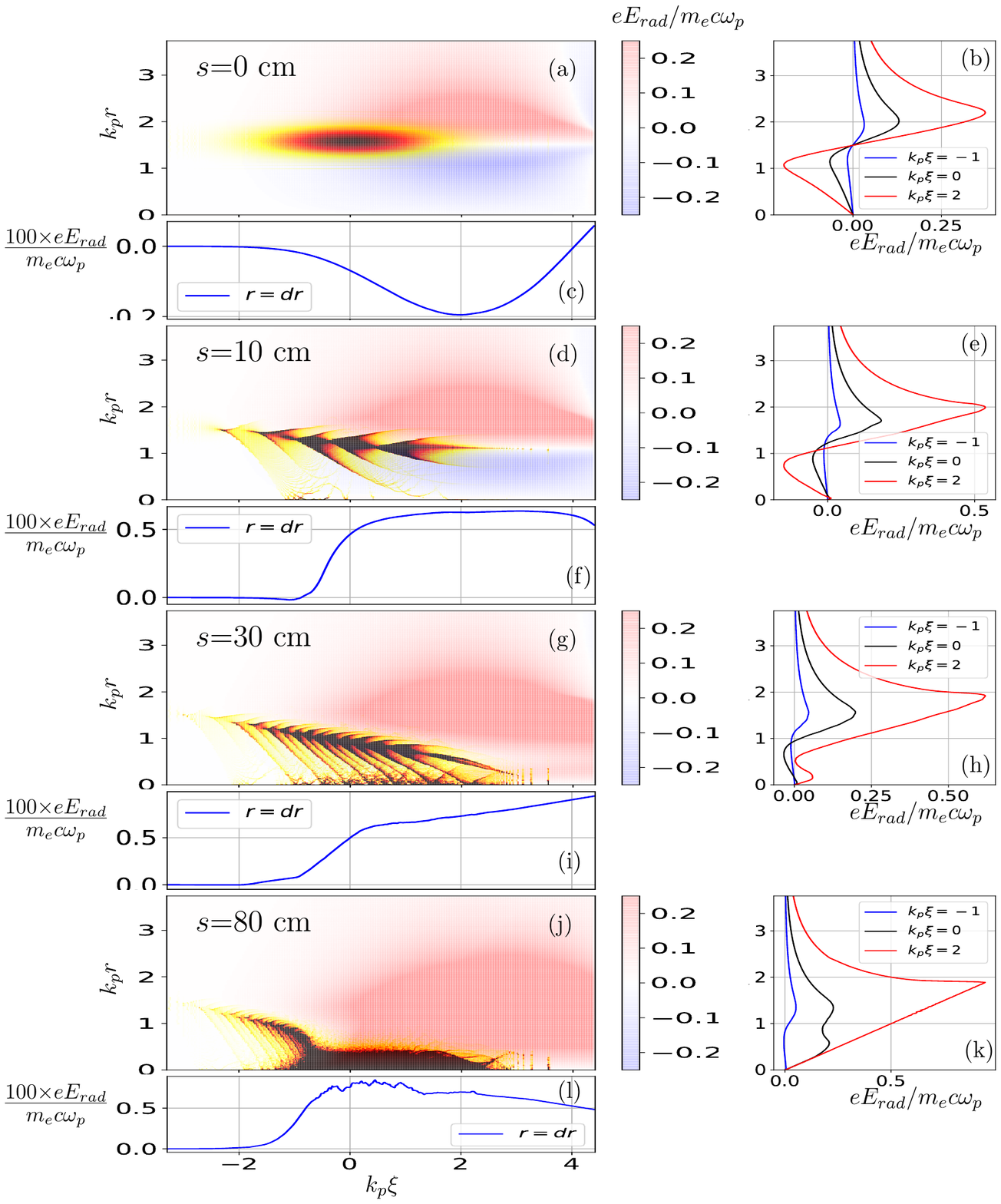}
\caption{Radial wakefield $eE_{rad}/m_ec\omega_p$ (in
  blue-white-red color map)   driven by a hollow electron beam of
  radius $k_pr_0=1.6$ moving towards left, at
  propagation distances $s$=0 cm (a), 10 cm (d), 30 cm (g) and
  80 cm (j).  Over-plotted is  the beam charge density $\rho_b/(n_{p0}e)$ in black-red-yellow color map saturated
  at $\rho_b/(n_{p0}e)$=-1. Radial and axial line-outs of $E_{rad}$ at
  a given $s$ are
  shown on the right of and below the corresponding color plot, respectively.\label{fig:erad_evolution_r08sr}}
\end{figure}

 Propagation of a hollow electron beam driver, 
 density profile given by  Eq. (\ref{eq:driver})  and shown in
Figs. \ref{fig:rhop_evolution_r08sr}a \& \ref{fig:erad_evolution_r08sr}a, in a plasma causes
 radial expulsion of the plasma electrons from the beam  towards and
 away from the beam axis at $r=0$ \cite{jain2015a}. Plasma ions pull back the expelled electrons
 which fall back behind the beam driver forming an azimuthally
 symmetric annular shape positively charged region in the moving
 computational domain, as shown in
 Fig. \ref{fig:rhop_evolution_r08sr}a. The positively charged annular region is
 bounded by closely spaced trajectories of plasma electrons, and thus
 plasma charge density $\rho_p$ is negative near (but outside) the radial
 boundaries of the annular region; the negative values being more
 pronounced towards the beam-rear as can be seen in
 radial line-outs of $\rho_p$ at $k_p\xi=-1$ (in the beam-front),
 $k_p\xi=0$ (beam center) and $k_p\xi=2$ (beam-rear) in
 Fig. \ref{fig:rhop_evolution_r08sr}b. The resulting radial wakefield  $E_{rad}=E_r-cB_{\theta}$
 changes direction across the bulk of the beam (Fig. 
\ref{fig:erad_evolution_r08sr}a and \ref{fig:erad_evolution_r08sr}b). Note that slope of a radial
line-out of $E_{rad}$, Fig. \ref{fig:erad_evolution_r08sr}b, at a radial
position where $E_{rad}=0$ increases towards beam-rear. 

Above the upper boundary, $\rho_p$
 reaches a negative peak value and then drops to vanish. Below the
 lower boundary, on the  other hand, $\rho_p$ tends to be radially
 uniform towards the beam axis after a small drop from its negative
 peak at the lower boundary. 
The difference in magnitude and radial variation of
 $\rho_p$ above and below the upper and lower boundaries,
 respectively, is because plasma electrons moving towards the beam axis
 through the beam's hollow region  repel each other, and thus are less
 radially deflected than those moving away from the axis.  
This makes the annular region  radially asymmetric with respect to the
beam center at $r=r_0$. Due to the radial asymmetry,
 the radial location of $E_{rad}=0$ with
respect to the beam center is slightly
shifted towards the beam axis (Fig. \ref{fig:erad_evolution_r08sr}a and \ref{fig:erad_evolution_r08sr}b).

\subsection{Development of fish-bone like structure in the  hollow 
  beam\label{sec:fishbone}}
Fig. \ref{fig:rhop_evolution_r08sr}  shows that the hollow beam driver, which initially had bi-Gaussian shape
(Fig. \ref{fig:rhop_evolution_r08sr}a), develops fish-bone like
structure by the time 
it propagates 10 cm into plasma (Fig. \ref{fig:rhop_evolution_r08sr}d). 
The structure develops as a result of
the $\xi$-dependent time period of   
the betatron oscillations of the beam electrons. The betatron
oscillations are caused by the
radial wakefield $E_{rad}$ which acts on beam electrons as a restoring
force  about the equilibrium positions where $E_{rad}$=0. 
The betatron time period $\tau_{\beta}$ under the influence of $E_{rad}$, whose
radial variation near the  equilibrium position in a $\xi$-slice
can be approximated as linear
(at least for first few tens of centimeters of propagation,
Figs. \ref{fig:erad_evolution_r08sr}b and \ref{fig:erad_evolution_r08sr}e), can be written as  
\begin{eqnarray}
\tau_{\beta}&=&2\pi\,\sqrt{\frac{m_e\gamma_b}{e\,[dE_{rad}/dr]_{E_{rad}=0}}}
\label{eq:betatron}
\end{eqnarray}
Here $[dE_{rad}/dr]_{E_{rad}=0}$ is the $\xi$-dependent slope of a radial line-out
of $E_{rad}$ at the equilibrium 
position,  
$\gamma_b$ the relativistic factor of the beam particles, $m_e$ 
the electron's rest mass and $e$ the electronic charge. For a uniform
region of ions, Gauss law gives $dE_{rad}/dr=n_{p0}e/2\epsilon_0$ and
Eq. (\ref{eq:betatron}) reduces to the usual expression
$\tau_{\beta}=\sqrt{2\,\gamma_b}\,\lambda_p/c$ for the betatron time period.

\begin{figure}
\includegraphics[clip=true,trim=2.5cm 12.0cm 2.5cm
  3.0cm,width=\linewidth]{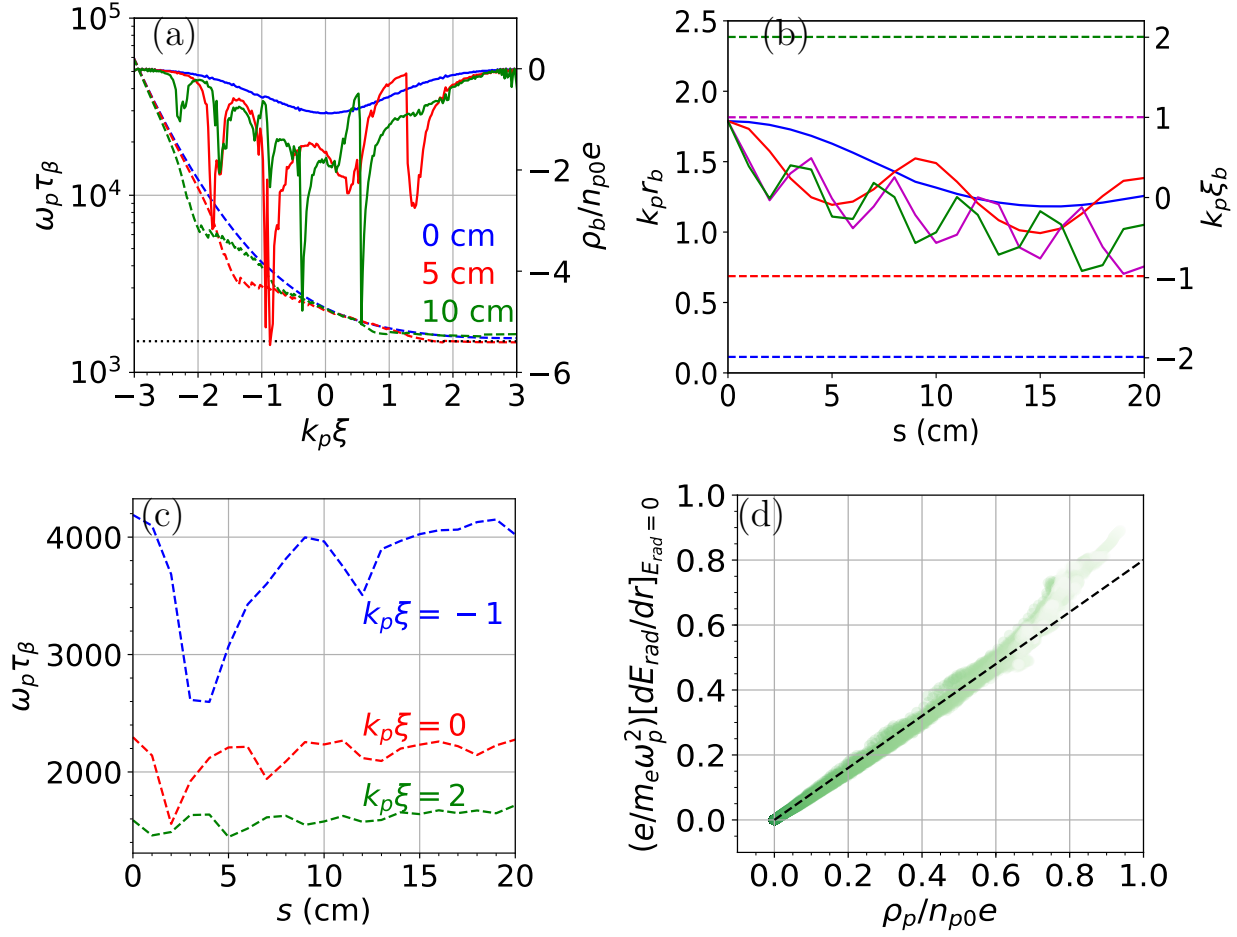}
\caption{In (a), betatron time period
  $\omega_p\tau_{\beta}$  (left axis, dashed lines)   calculated
  from Eq. (\ref{eq:betatron}) using   the initial value of $\gamma_b$
  and beam charge density $[\rho_b/(n_{p0}e)]_{E_{rad}=0}$ (right axis,
  solid lines) vs. $k_p\xi$ at $s$ = 0 cm (blue), 5 cm  (red) and 10 cm (green). A horizontal
  dotted line in (a) is for
  $\omega_{p}\tau_{\beta}=1.5\times10^3$. In (b), radial  ($k_pr_b$,
  left axis, solid lines) and axial ($k_p\xi_b$, right axis, dashed
  lines) positions vs.   $s$  for select   beam particles with
  same initial radial $r\approx  r_0+\sigma_r=1.8 k_p^{-1}$ but
  different axial positions $k_p\xi$=-2  (blue), -1 (red),  1 (magenta)
  and 2 (green). In (c), $\omega_{p}\tau_{\beta}$ vs. $s$ at $k_p\xi=-1$ (blue), $k_p\xi=0$ (red) and
  $k_p\xi=2$ (green).   
  In (d), scatter plot $(e/m_e\omega_p^2)[dE_{rad}/dr]_{E_{rad}=0}$ vs. 
  $[\rho_p/(n_{p0}e)]_{E_{rad}=0}$ for $k_p\xi < 3$ (green color
  representing $\xi$-location fades towards the beam-rear) and $s\leq$
  20 cm, and a fit
  $(e/m_e\omega_p^2)[dE_{rad}/dr]_{E_{rad}=0}=0.8\,[\rho_p/(n_{p0}e)]_{E_{rad}=0}$
  (dashed line).  
\label{fig:taubeta}}
\end{figure}

Fig. \ref{fig:taubeta}a shows dependence of $\tau_{\beta}$ on $\xi$ at different propagation
distances $s$. We used initial value of $\gamma_b$ in
Eq. (\ref{eq:betatron}) for the calculation
of $\tau_{\beta}$ as the axial wake field $E_z \sim$ 5 GV/m  (not
shown here)   decelerating the beam reduces the initial beam energy of
23 GeV by a small amount $\sim eE_z\times$ 10 cm=0.5 GeV in first 10 cm
of propagation. 
By the time $\Delta t_{10\mathrm{cm}}=10\,
\mathrm{cm}/c\approx 4.2\times 10^3/\omega_{p}$ the beam propagated 10 cm in the
plasma developing fish-bone like structure, average change in
$\tau_{\beta}$ from its initial value (at $s$=0 cm) is relatively
small (Fig. \ref{fig:taubeta}a). On the other hand, $\tau_{\beta}$ changes by an order of
magnitude from the front to the rear of the beam. It drops from its value
$\omega_{p}\tau_{\beta}^{front}\approx 10^4$ in the beam front
($k_p\xi=-2$) to reach an asymptotic  value
$\omega_{p}\tau_{\beta}^{rear}=1.5\times 10^3$ in the beam rear
($k_p\xi=2$). 
Consequently, the beam electrons in
different $\xi$-slices  
 will be in different phases of the oscillations at  a given
 propagation distance $s$ as shown in Fig. \ref{fig:taubeta}b. 
For example, by the time $\Delta t_{10\mathrm{cm}}$,  the electrons in
the beam-rear with their initial axial positions at $k_p\xi=1$
and 2, for which values of $\tau_{\beta}$ are only slightly different, are about to complete their third oscillation ($\Delta
t_{10\mathrm{cm}}/\tau_{\beta}^{rear}\approx 2.8$) while  those in the
beam-front with initial axial position at $k_p\xi=-2$ are only approximately halfway of their first oscillation ($\Delta
t_{10\mathrm{cm}}/\tau_{\beta}^{front}\approx 0.42$).

Since the beam electrons in neighboring $\xi$-slices
are in different phases of their radial oscillations, the beam
density, which at a point ($r$,$\xi$) has contributions from
neighboring electrons including those in the neighboring $\xi$-slices,
develops a structure different from its initial bi-Gaussian
structure.
 The new structure would depend on the nature of the
variation of phases with $\xi$.  In the beam-rear where the $\xi$-dependence
of $\tau_{\beta}$ and thus of oscillation phase is relatively flat,
the beam electrons, for example those with the initial axial positions
at $k_p\xi=1$ and $k_p\xi=2$ shown in Fig. \ref{fig:taubeta}b, oscillate
with almost similar phases (at least up to $s$=10 cm) resulting in a beam-rear 
relatively  stretched in $\xi$. On the other hand, faster variation of
$\tau_{\beta}$ with $\xi$ towards the beam-front results in the
development of fish-bone like structure in which wings separated by
intervals of pinched beam develop from the front to the rear of the
beam. The pinched parts of the beam appears as multiple peaks in beam charge density shown in Fig. \ref{fig:taubeta}a. Note that the beam electrons
do not significantly change their $\xi$-positions during the beam
propagation, as can
be seen from Fig. \ref{fig:taubeta}b. Thus moving of the beam
electrons to the neighboring slice is not the reason for the fish-bone
like structure as may appear from the
wings of the structure.

After propagating a finite distance in plasma, the peaks in beam charge density, Fig. \ref{fig:taubeta}a, are enhanced over
the initial single peak value. The first
enhanced peak in  the beam front, for example at $k_p\xi\approx -1.75$
at $s$=5 cm, causes much stronger expulsion of the plasma electrons
from the beam path than that by the initial beam density, 
augmenting the positive charge density of plasma behind the peak due
to the enhanced exposition of the background ions. As a result,
$[dE_{rad}/dr]_{E_{rad}=0}$, which has positive correlation and for
$k_p\xi < 2$ 
linear scaling with $[\rho_p]_{E_{rad}=0}$    as 
$(e/m_e\omega_p^2)[dE_{rad}/dr]_{E_{rad}=0}=0.8\,[\rho_p/n_{p0}e]_{E_{rad}=0}$
(Fig. \ref{fig:taubeta}d) --- a reminiscent
of the formula $dE_{rad}/dr=\rho_p/2\epsilon_0$ for a uniform
background of ions, increases. This cause a local 
drop in the value of $\tau_{\beta}$ according to
Eq. (\ref{eq:betatron}), which appears behind the first enhanced peak 
as an ankle-like shape in the $\tau_{\beta}$-vs.-$\xi$ curves (Fig. \ref{fig:taubeta}a). 
Since beam density in a 
 given $\xi$-slice  
oscillates with $s$, the first peak at $s$=5 cm subsides and a new
first peak at $k_p\xi\approx -2.3$ with a corresponding ankle in
$\tau_{\beta}$-vs.-$\xi$ curve behind it forms  by
$s$=10 cm.
The value of $\tau_{\beta}$ drops at other $\xi$-locations as
well but by small amount and does not seem to be associated with the peaks
of the beam density.    
 In fact, $\tau_{\beta}$ at a given $\xi$ 
oscillates with $s$ in such a way that its  maximum value does not
exceed its initial value, Fig. \ref{fig:taubeta}c. 


\subsection{Differential shifting of the beam's bulk towards its
  axis \label{sec:bending}}
\begin{figure}
\includegraphics[clip=true,trim=2.5cm 6.5cm 2.5cm
  2.5cm,width=\linewidth]{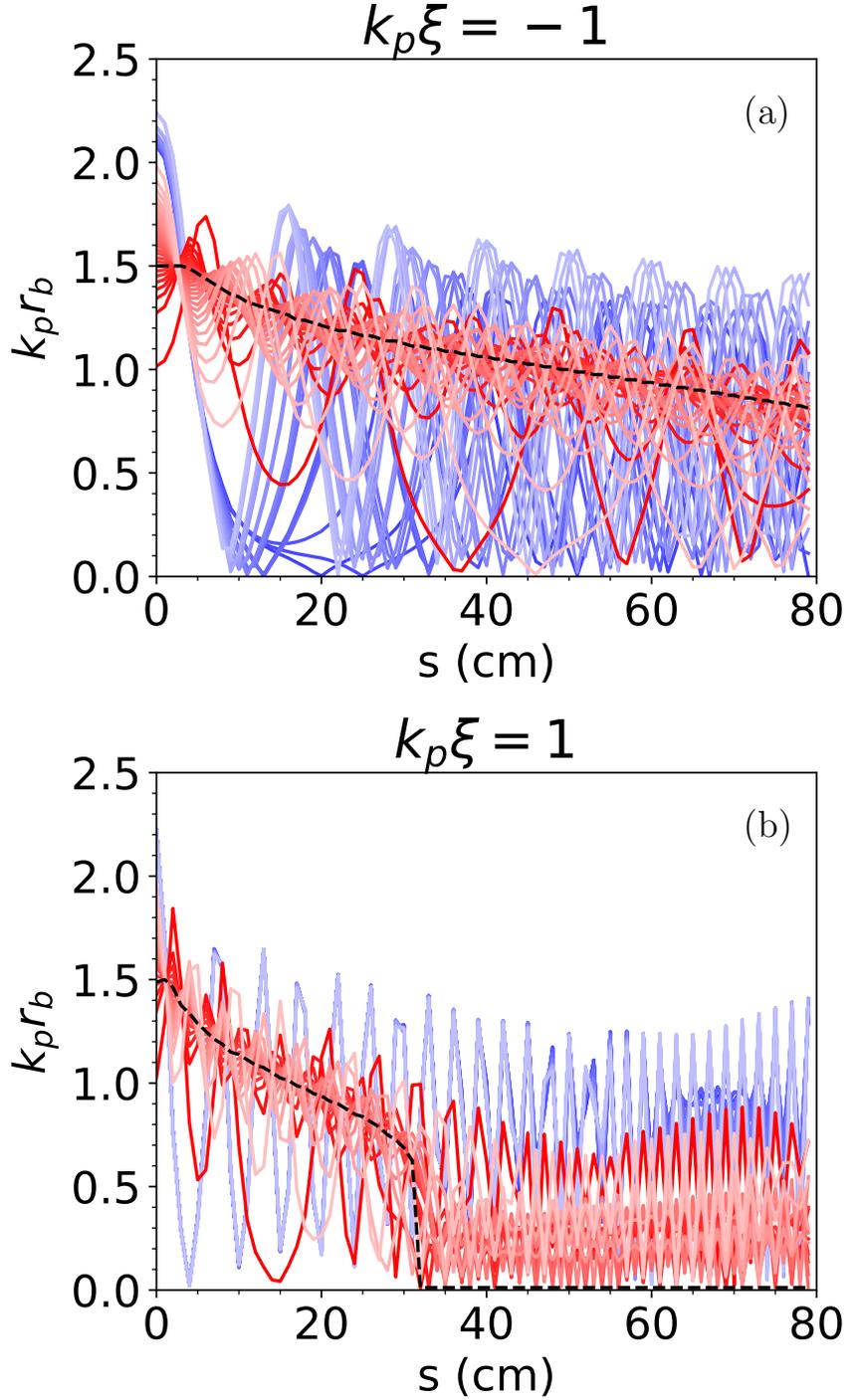}
\caption{Radial positions $r_b$ (solid lines) of select   beam
  particles and equilibrium position $r_{eq}$ (dashed line) in the
  $\xi$-slices $k_p\xi=-1$ (a) and 1 (b) as a function of the distance
  $s$ traveled by the  beam. Lines with shades of blue (red) represent
  the particles whose   initial radial positions are in the region
  where $-\psi >0 $ ($-\psi < 0$). Shades are used to distinguish
  different particles represented by a same color.  
  \label{fig:rb_psiLt0_psiGt0}}
\end{figure}

Simultaneous with the development of fish-bone structure, the hollow beam
shifts its bulk, rear part fastest, towards the beam  axis as shown in Fig. \ref{fig:rhop_evolution_r08sr}. 
The shift results from the betatron oscillations of the beam
electrons about an equilibrium position 
($E_{rad}=0$) which shifts radially towards the beam axis
during beam propagation. Fig. \ref{fig:rb_psiLt0_psiGt0}
 shows radial oscillations of select
beam particles and the changing radial position $r_{eq}$ of the equilibrium  at
locations in the beam front 
($k_p\xi=-1$) and beam rear ($k_p\xi=1$) as a function of distance $s$ traveled by the beam. 
At $s=0$, equilibrium position $k_pr_{eq,0}=1.5$ is already radially
shifted towards the beam axis with respect to the peak beam density at $k_pr=k_pr_0=1.6$.
  This shift is same for all $\xi$-coordinates as can be noticed in Fig. \ref{fig:erad_evolution_r08sr}a.
 
As beam begins to propagate in plasma, beam electrons in a given
$\xi$-slice 
 oscillate about the initial shifted equilibrium position  to reach
 the equilibrium simultaneously, and thus form a density peak there, after
 approximately a quarter of their first  betatron oscillation in that
 $\xi$-slice.  The density peak forms faster in the rear part of the
 beam as the betatron oscillations are faster there. Consequently, initially
 hollow beam takes an overall shape of a conical frustum with the
 radius decreasing from front to the rear of the beam.
Since the effective beam radii for all $\xi$ are now less than the
initial beam radius $r_0$,  propagation of now frustum-shape
beam drives wake fields with the equilibrium position further shifted towards the beam
axis. Indeed, Fig. \ref{fig:rb_psiLt0_psiGt0} shows that the equilibrium
position begins to shift towards the beam axis  after approximately a
quarter of the first betatron oscillations in a given $\xi$-slice. As
the frustum shape beam propagates further, beam electrons adjust to oscillate
about the new equilibrium position forming the density peak there  and the process continues until the beam
reaches too close to the axis.
In this way, bulk of the beam shifts  towards the axis. 


Note that only the beam electrons close to the equilibrium position
performs betatron oscillation 
about the changing equilibrium position.  This is because $E_{rad}$ can be approximated as linear only
in the neighborhood of the equilibrium position. Away from the
equilibrium positions, $E_{rad}$ is nonlinear in $r$ and particle
motion is no longer a simple betatron motion.

\subsection{Collapse of the hollow beam into its axis\label{sec:collapse}}
\begin{figure}
\includegraphics[clip=true,trim=1.55cm 5.8cm 3.45cm
  2.5cm,width=\linewidth]{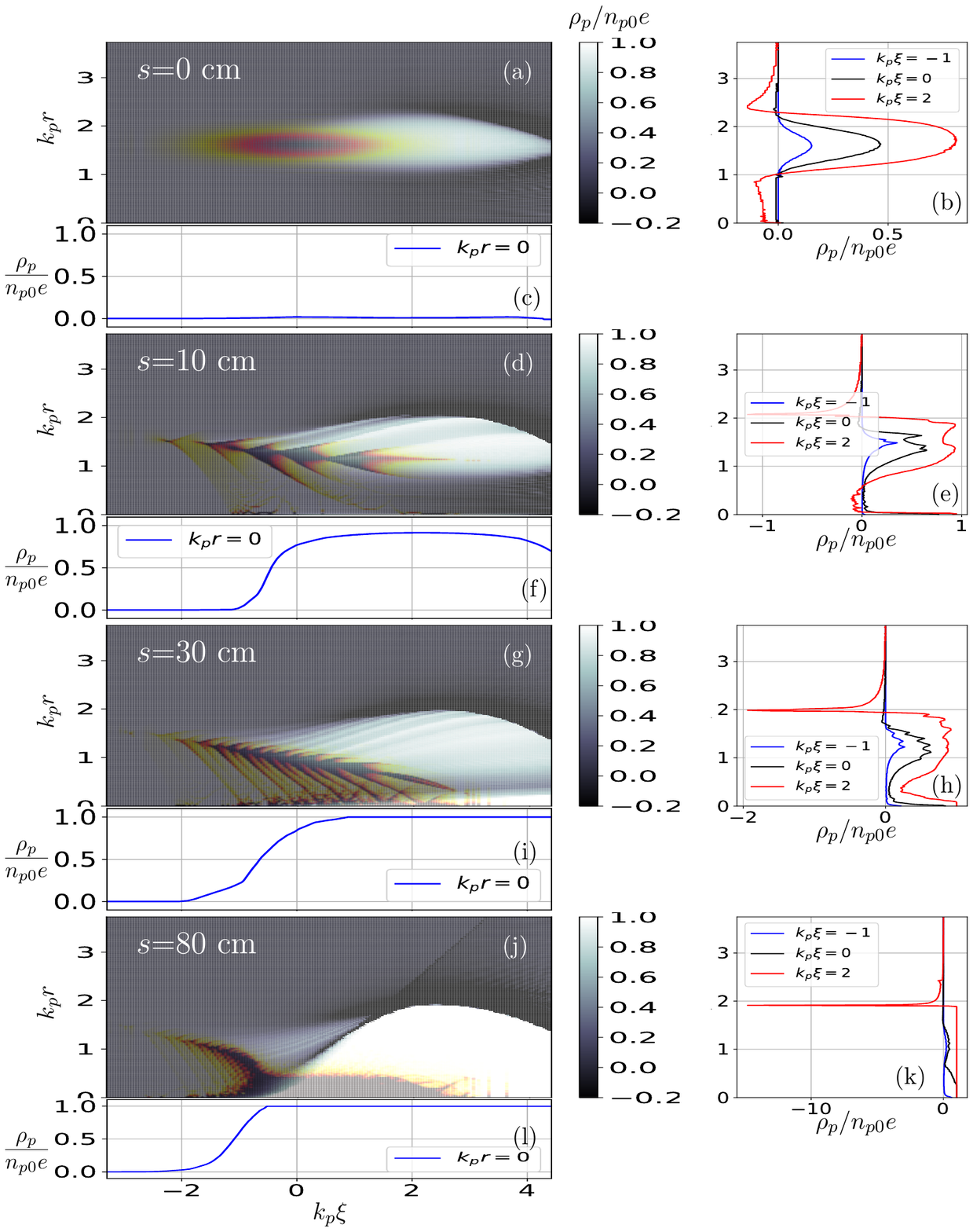}
\caption{Plasma charge density $\rho_p/(n_{p0}e)$ (in
  black-white color map)   driven by a hollow electron beam of radius 
  $k_pr_0=1.6$ moving towards left, at
  propagation distances $s$=0 cm (a), 10 cm (d), 30 cm (g) and
  80 cm (j).  Over-plotted is  the beam charge density $\rho_b/(n_{p0}e)$ in black-red-yellow color map saturated
  at $\rho_b/(n_{p0}e)$=-1. Radial and axial line-outs of $\rho_p$ at
  a given $s$ are
  shown on the right of and below the corresponding color plot, respectively.\label{fig:rhop_evolution_r08sr}}
\end{figure}

Figs. \ref{fig:rhop_evolution_r08sr}  \& \ref{fig:erad_evolution_r08sr}  
show that the beam finally collapses into its axis. The reason for
the collapse is the interaction of the beam electrons with the wakefields generated near the beam axis by some of the beam particles
reaching the axis.
Fig. \ref{fig:rhop_evolution_r08sr}d or
\ref{fig:erad_evolution_r08sr}d shows that  by $s=10$ cm beam density has already
got accumulated   on the axis  mainly in the region $-1.5
< k_p\xi < 0$. This accumulation is due to the beam electrons reaching the
axis in the course of their motion under the influence of the radial
force,   $F_{rad}=-e\,E_{rad}=-e\,\partial
(-\psi)/\partial r$, where $\psi(r,\xi,t)=\phi-cA_z$ is the space and
time dependent wake field potential. Energy of the radial electron
motion,
\begin{eqnarray}
W_r=\frac{\gamma_bm_ev_r^2}{2}+e(-\psi)
\end{eqnarray} 
can be shown to be approximately constant under the assumption of slow
time variations of $\gamma_b$, $\psi$ and electron's axial position
$\xi_b$. These assumptions are met for a few tens of centimeters of
beam propagation.

\begin{figure}
\includegraphics[clip=true,trim=2.0cm 12.5cm 3.5cm
  2.5cm,width=\linewidth]{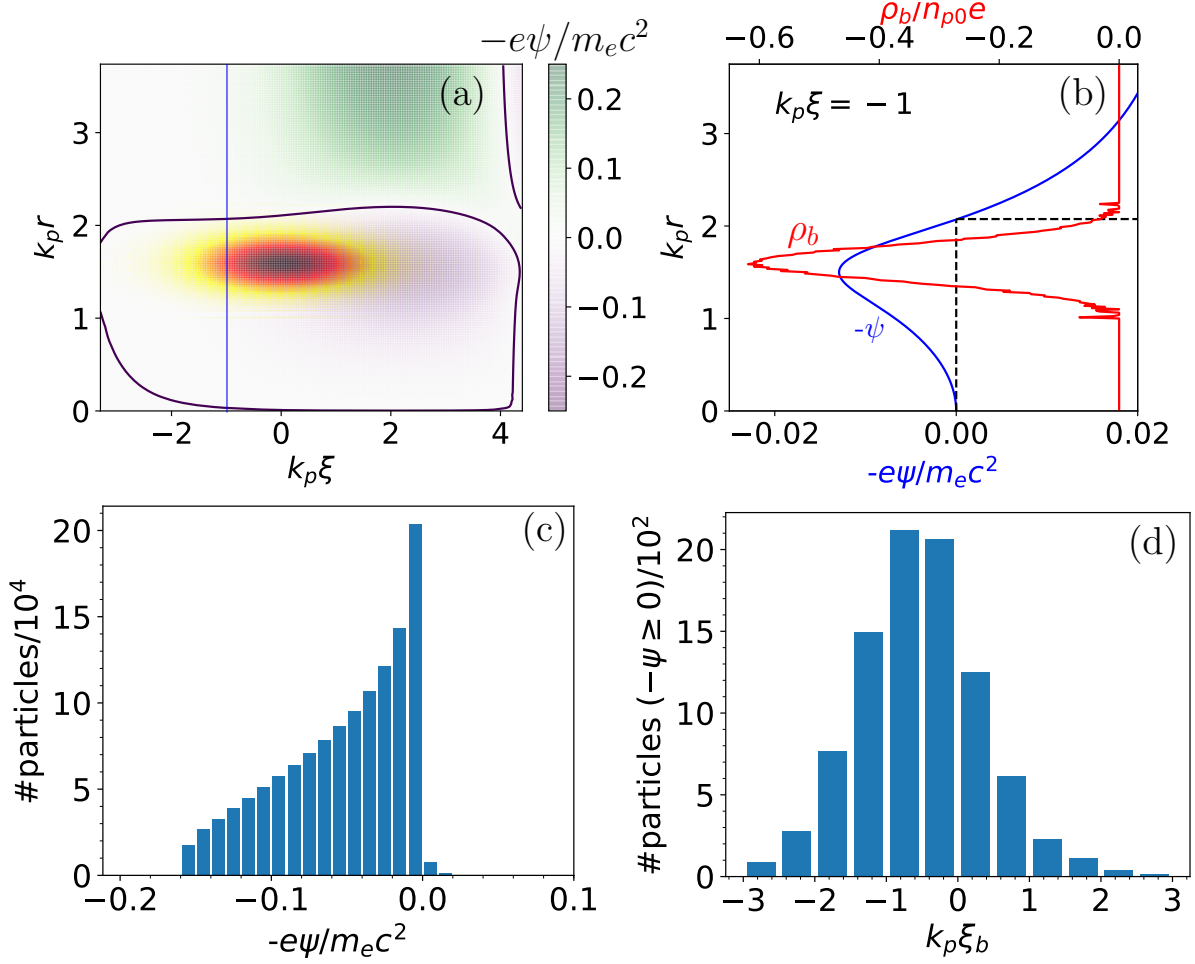}
\caption{Negative of wakefield potential, i.e., $-e\psi/m_ec^2$,
  (purple-green color map), a contour  of $-e\psi/m_ec^2=10^{-5}$ (black line)
  and $\rho_b/(n_{p0}e)$ (red-yellow color map) at $s=0$ (a).  Radial line-outs of $\rho_b/(n_{p0}e)$ (red, top horizontal axis) and
  -$e\psi/m_ec^2$ (blue, bottom horizontal axis) along the blue vertical line
  drawn in (a) 
  at $k_p\xi$= -1 (b). Horizontal and vertical dashed lines in (b) meet the $\psi$-curve where $\psi=0$.
Histogram of the values of $-e\psi/m_ec^2$ at particles
  position (c). Histogram of the $\xi$-positions of the particles for
  which $-\psi \geq 0$ (d).  \label{fig:psi}}
\end{figure}

Figs. \ref{fig:psi}a and  \ref{fig:psi}b show negative of the wake
field potential ($-\psi$) at $s=0$. Since 
$\psi=0$ on the beam axis and initial radial velocities of the beam
electrons are zero,
only the beam electrons at initial radial positions  where $-\psi \ge
0$ can reach the axis. 
On the other hand, the beam electrons
with positive values of $\psi$ significantly different from zero will
not reach the axis.  
Fig. \ref{fig:psi}c shows that there are small
 but finite number of particles for which $-\psi \geq 0$.
 Histogram of the $\xi$-positions of these  particles,
 Fig. \ref{fig:psi}d, shows that they are distributed along the whole
 length of the beam 
 from its front to rear but are mainly concentrated at the axial
 positions in the range $-1.5 < k_p\xi <0$ where beam density first
 accumulates on the axis. 
 Contour of a very small value of 
$-e\psi/m_ec^2 =10^{-5}$, which  is also closest to the beam in the region $-1.5
< k_p\xi < 0$, show that the particles with $-\psi \ge 0$ are located
at the radially outer edge of the beam (Fig. \ref{fig:psi}a). 
These particles account for small but finite  beam  density at the
outer beam edge where 
$|\psi| \approx 0$, as illustrated by  radial line-outs of $\psi$ and
$\rho_b$ at $k_p\xi=-1$ in  Fig. \ref{fig:psi}b.

Fig. \ref{fig:rb_psiLt0_psiGt0} shows that the electrons which have
their initial radial positions in the region where $-\psi > 0$ indeed reach
close to the axis in the course of their first oscillation. The
particles at $k_p\xi=-1$ reach the axis first time around $s$=10 cm later  than
those at $k_p\xi=1$ due to their slower oscillations. Since the particles at $k_p\xi=-1$ are larger in number and spend
sufficient time  $\Delta t=1 \mathrm{cm}/c\approx 420 \,
\omega_{pe}^{-1}$ near the axis, the beam density accumulated on the
axis at  $k_p\xi=-1$ is larger than  that at
$k_p\xi=1$. The on-axis beam density at $k_p\xi=-1$ is
also larger than the background plasma density and thus excite secondary wake
fields near the axis  similar to those excited by a solid electron
beam in the blow-out regime, as can be seen by 
 axial line-outs of plasma density and radial wake field in Figs. \ref{fig:rhop_evolution_r08sr}f and
\ref{fig:erad_evolution_r08sr}f. At $s$=10 cm,  
$\rho_p$ and $E_{rad}$ on the axis are positive over an axial length
of the order of a plasma wavelength $2\pi/k_p$ starting around
$k_p\xi=-1$. They are not, however, well extended in the radial
direction at $s$=10 cm. With the further propagation of the
beam, the secondary wake gets radially extended 
(Figs. \ref{fig:rhop_evolution_r08sr}h and  \ref{fig:erad_evolution_r08sr}h). 
By $s$=30 cm, beam electrons, which were no reaching the
axis, i.e., those with the initial position where $-\psi < 0$,  begin
to access the radially extended 
secondary wake
field, first in the rear part of the beam, Figs. \ref{fig:rhop_evolution_r08sr}g and
\ref{fig:erad_evolution_r08sr}g. These electrons are pulled by the
secondary radial wake field to trap them near the axis, as shown in
Fig. \ref{fig:rb_psiLt0_psiGt0}b.  This process continues from the
beam-rear to beam-front finally collapsing a significant part of the beam into the axis.

\section{Dependence on hollow beam radius \label{sec:dependence_r0}}
\begin{figure}
\includegraphics[clip=true,trim=0.0cm 3.75cm 0.0cm 0.5cm,width=\linewidth]{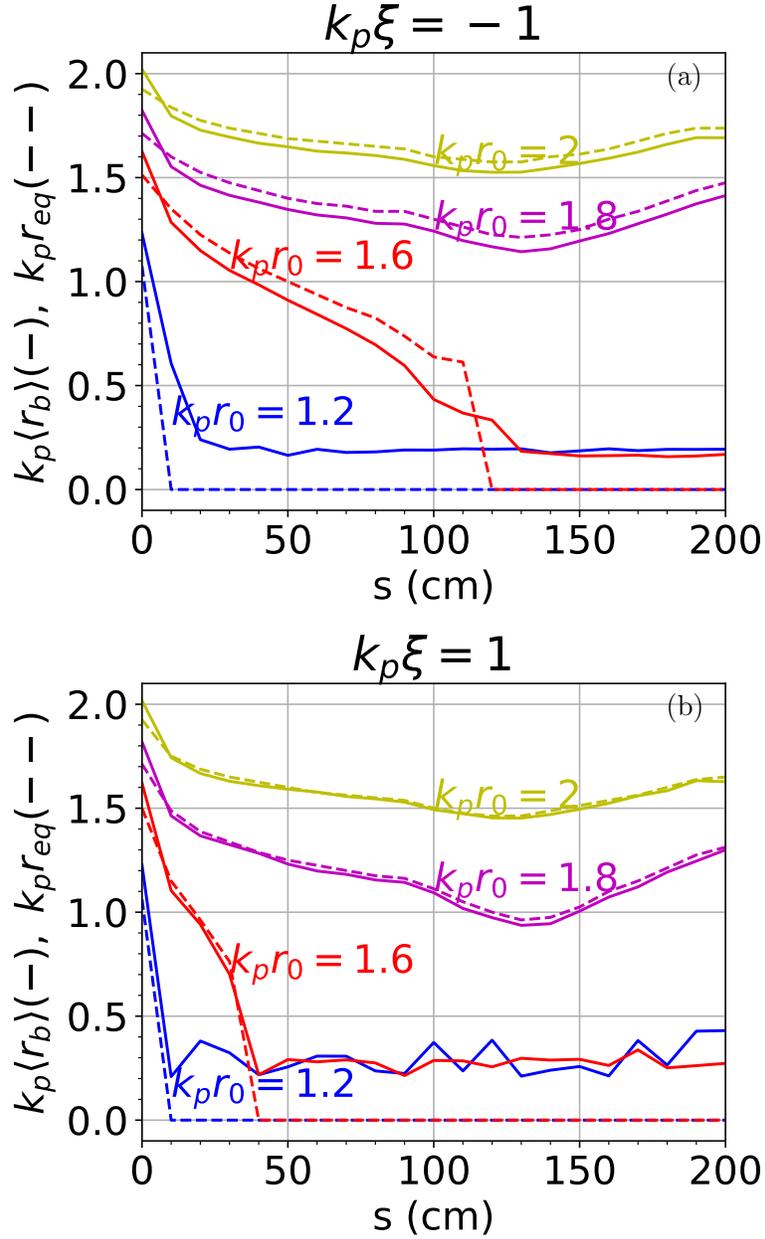}
\caption{Mean radial position $\langle r_b \rangle$ of beam electrons (solid line) and radial position $r_{eq}$ of the equilibrium (dashed line) at $k_p\xi=-1$ (a) and $k_p\xi=1$ (b) as a function of the distance
  $s$ traveled by the  beam for various value of $r_0$. 
  \label{fig:mean_rb_req}}
\end{figure}

\begin{figure}
\includegraphics[clip=true,trim=1.15cm 7.5cm 0.5cm
  1.6cm,width=\linewidth]{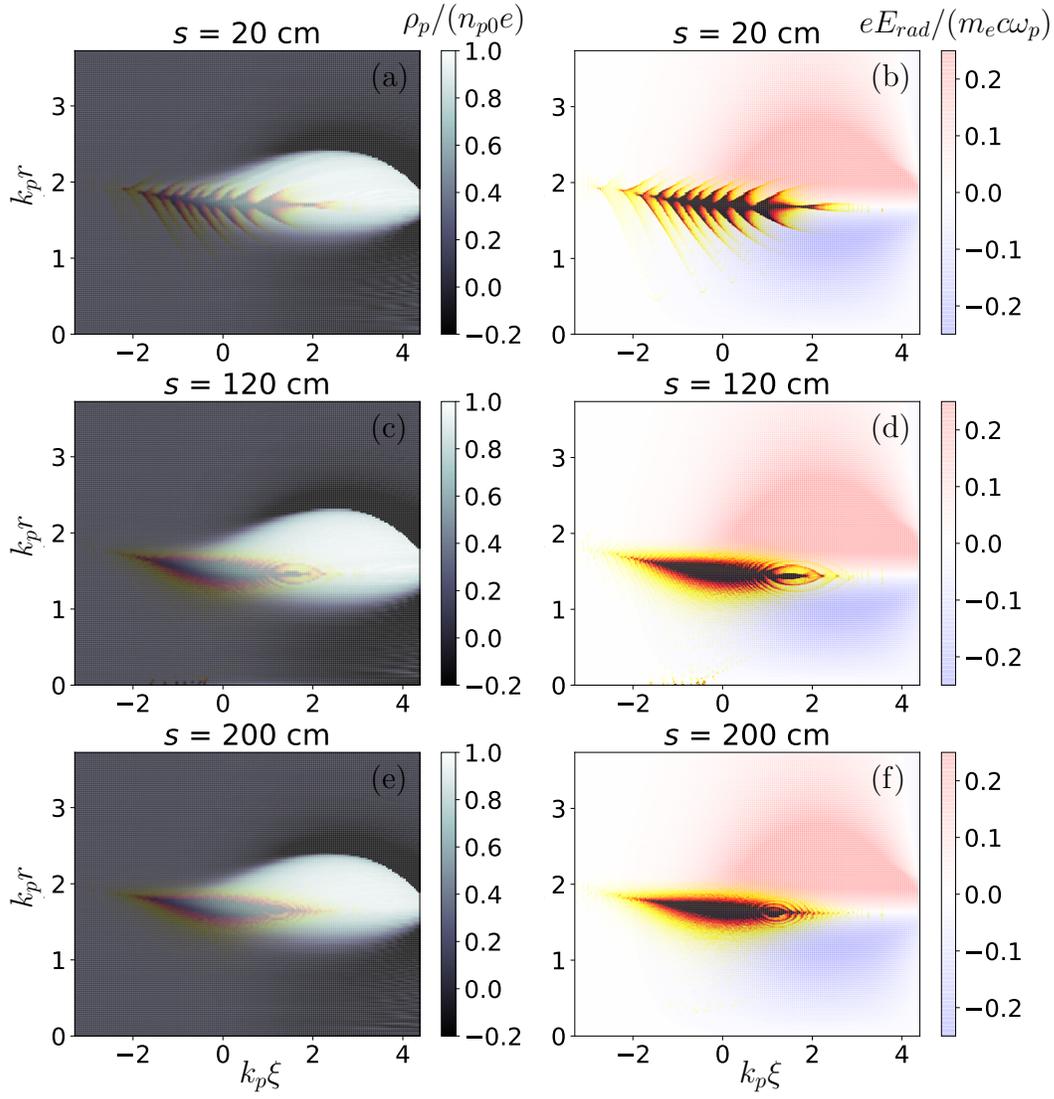}
\caption{
Plasma charge density $\rho_p/(n_{p0}e)$ (left column, black-white color map) and radial wakefield $e E_{rad}/m_e c \omega_{pe}$ (right column, blue-white-red
color map) driven by a hollow electron beam ($n_{b0} = n_p$ ,
$k_pr_0 = 2.0$) moving towards left, at various propagation distances. Over-plotted is
the beam charge density $\rho_b/(n_{p0}e)$ in black-red-yellow color map saturated at $\rho_b /(n_{p0} e)=-1$.
 \label{fig:evolution_r010sr}}
\end{figure}

Fig. \ref{fig:mean_rb_req} shows 
mean radial position $\langle r_b \rangle$ of beam electrons, defined  at $\xi=\xi_0$ as,
\begin{eqnarray*}
\langle r_b \rangle &=&\frac{1}{N}\sum_{i=1}^{N} r_{b,i}\,\, ,\,\,
\forall \,\, i: \xi_0-d\xi/2 \le \xi_{b,i} \le \xi_0+d\xi/2    
\end{eqnarray*}
 and radial equilibrium position $r_{eq}$ for $k_p\xi_0=-1,\,\, 1$ and various values of $r_0$.  
 Hollow electron  beam shifts towards (represented by drops in $\langle r_b
 \rangle$ and $r_{eq}$ from their initial value)   and collapses into
 its axis ( represented by $r_{eq}$ attaining zero value), starting
 from its rear (Fig. \ref{fig:mean_rb_req}b) and progressing towards front
 (Fig. \ref{fig:mean_rb_req}a),  for $k_pr_0=1.2$ and 1.6. For larger
 radii, $k_pr_0=1.8$ and 2.0, hollow beam does shift towards its axis but  not  collapse
 even after propagating 200 cms in plasma. Instead, $\langle r_b
 \rangle$ and $r_{eq}$ simultaneously stop dropping and rise again
 after a certain propagation distance, e.g., at approximately 120 cm
 for $k_pr_0=2.0$. The rate of drop,  distance at which the drop
 stops and the amount of drop in $\langle r_b \rangle$ and
 $r_{eq}$ decrease with the beam radius, resulting in a stable propagation of hollow beam in
 plasma for large radius, as shown in Fig. \ref{fig:evolution_r010sr} for
 $k_pr_0=2.0$.  

\begin{figure}
\includegraphics[clip=true,trim=2.15cm 13.0cm 1.5cm
  1.6cm,width=\linewidth]{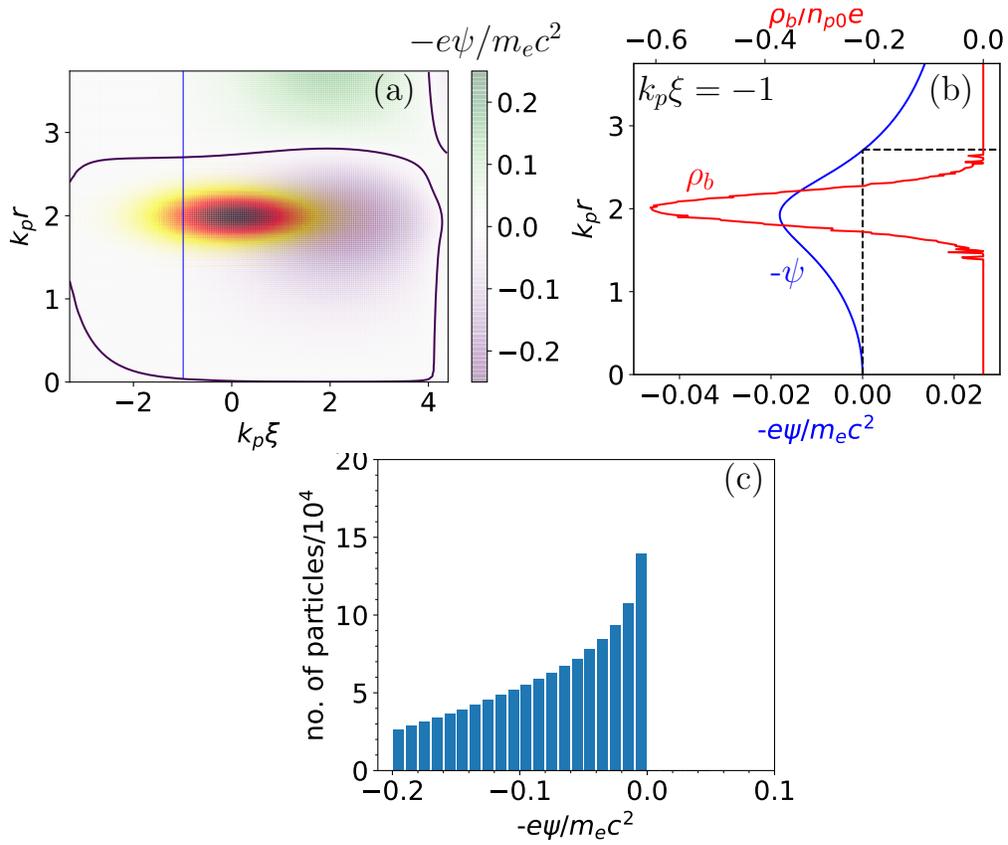}
\caption{(a), (b) and (c) are the same as those in Fig. \ref{fig:psi}
  but here 
  for $k_pr_0=2.0$.
\label{fig:psi_r010sr}}
\end{figure}

 The reason of no beam collapse for large beam radius can be
 understood from  Fig. \ref{fig:psi_r010sr} which shows negative of the wake field potential ($-\psi$) at $s=0$ for $k_pr_0=2.0$. The contour of $-e\psi/m_ec^2=10^{-5}$ in Fig. \ref{fig:psi_r010sr}a is fairly away from the beam edges so that 
the beam density vanishes at the locations where $|\psi| = 0$ 
(Fig. \ref{fig:psi_r010sr}b). Consequently, there are no beam
electrons with $-\psi > 0$, as shown in Fig. \ref{fig:psi_r010sr}c, to
reach and excite secondary wakefields on the axis, avoiding the collapse.  

Fig. \ref{fig:evolution_r010sr}c and \ref{fig:evolution_r010sr}d shows
 that a small amount of beam density does accumulate near the axis after propagating a distance 
 $s\sim $ 100 cm,  much later in comparison to the case
 of $k_pr_0=1.6$. This accumulation of beam density is, however, not
 due to the beam electrons with $-\psi_{s=0} >
 0$ as in the case of $k_pr_0=1.6$, but rather due to the 
 evolution of $\psi$ during beam propagation such that some beam
 electrons have $-\psi >0$. It, however, disappears by $s=200$ cm,
 Figs. \ref{fig:evolution_r010sr}e and \ref{fig:evolution_r010sr}f,
 without affecting the beam propagation. Therefore, number of beam
 electrons with $-\psi >0 $ at $s=0$ seems to control the collapse of
 the   beam.
\section{Conclusion \label{sec:conclusion}}
We have shown by quasi-static kinetic simulations using the code WAKE  \cite{jain2015a}
that an ultra-relativistic  hollow 
electron beam propagates in a plasma by developing fish-bone like
structure and shifting its bulk, differentially along its length (rear part
fastest), towards its axis.  
The fish-bone structure develops because electrons performing betatron
oscillations in the neighboring beam cross-sections are not in-phase
due to the continuous drop of the betatron time period from front to the
rear part of the beam.
Beam shifts towards its axis due to betatron oscillations of the beam electrons
about an equilibrium position which shifts radially towards the beam axis
during the beam propagation. The rear part shifts fastest due to the
faster betatron oscillations there. For small beam radius, rear part
of the beam
shifting towards the beam axis is relatively quickly pulled towards the
beam axis by the secondary wake fields generated by some of the beam electrons
reaching the axis.
As a result, hollow beam 
collapses  into its axis with the collapse progressing from its rear
to front part. 
For beams with sufficiently large radius, not many particles reach the
axis and the beam can thus propagate stably in plasmas for several meters.  For
the parameters in our simulations, beam with radius $k_pr_0 \geq 1.8$ propagates
stably up to 200 cms in plasma.   

Propagation of ultra-relativistic hollow electron beams in plasmas drive 
wakefields suitable for positron acceleration (radial field focusing,
axial field accelerating) near the beam axis  \cite{jain2015a}. Collapse of the beam into its
axis causes the radial wake field near the axis to be
defocusing for positrons and thus destroys the wakefield structure  for
positron acceleration.  
Hollow electron beams with sufficiently large
radius do not collapse but shifts radially towards their axis. The
radial shift, however, does not change the wake field structure for
positron acceleration.  A peak positron accelerating wake field of
2.5 GV/m is produced at the beam axis in our simulations of stable beam
propagation with
$k_pr_0=2.0$. This accelerating field is of the same order of
magnitude as observed in experiments  of positron
acceleration by propagation of positron bunch in 
hollow plasma channels \cite{corde2015,doche2017}. Moreover,
accelerating field scales linearly with the total charge in
the hollow beam driver. Thus,
positron acceleration by plasma wakefields driven by hollow electron
beams is an attractive scheme for its future application in
electron-positron collider.




\appendix*

\begin{acknowledgments}
 Author thanks T. M. Antonsen, Jr. (University of Maryland,
 College Park, MD, USA) for providing the code WAKE. Author 
 thanks T. M. Antonsen and J. P. Palastro (University of
 Rochester, New York, USA) for initial discussions on the subject.       
\end{acknowledgments}

%


\end{document}